# Nonlinear imaging and 3D-mapping of terahertz fields with Kerr media

Matteo Clerici,[1,2,*] Daniele Faccio,[2] Lucia Caspani,[1] Marco Peccianti,[3] Eleonora Rubino,[4] Luca Razzari,[1] François Légaré,[1] Tsuneyuki Ozaki,[1] and Roberto Morandotti[1]

[1]INRS-EMT, 1650 Blvd. Lionel-Boulet, Varennes, Québec J3X 1S2, Canada.
[2]School of Engineering and Physical Sciences, Heriot-Watt University, SUPA, Edinburgh EH14 4AS, UK.
[3]Institute for Complex Systems (ISC), CNR, via dei Taurini 19, 00185 Rome, Italy.
[4]Dipartimento di Scienza e Alta Tecnologia, Università degli Studi dell'Insubria, via Valleggio 11, 22100 Como, Italy.

We investigate the spatially and temporally resolved four-wave mixing of terahertz fields and optical pulses in large band-gap dielectrics, such as diamond. We show that it is possible to perform beam profiling and space-time resolved mapping of terahertz fields with sub-wavelength THz resolution by encoding the spatial information into an optical signal, which can then be recorded by a standard CCD camera.

Since the seminal work of Hu and Nuss [1], time-resolved imaging of terahertz (THz) radiation (T-Ray imaging) has become a widely investigated and employed technique. The initial efforts in the nineties (see e.g. [1–4] and references therein) triggered the interest of the optics community, and a large literature followed (see e.g. [5,6] for a review). Currently, T-Ray imaging systems are commercially available and widely employed for a diverse range of applications, both in research labs and for applied purposes. For several applications however, spectroscopic information is not required and hence imaging techniques relying on the analysis of the spatially resolved electro-optical effect in second order crystals may result in an unrequired effort. For such cases, THz focal plane detectors are the most straightforward tools. Although research on THz focal plane arrays is proceeding very rapidly (see e.g. [7,8]), such devices are still not readily available or as efficient as CCD and CMOS cameras for the visible part of the spectrum and are typically quite expensive.

Alternatively, the detection of THz fields can also be performed via the THz electric-field-induced second harmonic generation in Kerr media, namely EFISH (or TFISH, when considering the THz electric field). Such an effect is a degenerate case of Four-Wave Mixing (FWM) between an unknown THz and an optical probe field, which up-shifts the frequency of the THz pulse into the visible, where it could be easily recorded with standard silicon detectors as a function of the delay with the probe pulse. This effect has been employed e.g. for the detection of freely propagating millimeter waves and THz pulses [9,10]. Furthermore, TFISH in gas media is the key for the coherent detection of ultra-broadband THz pulses [11,12].

In this Letter, we show that spatially resolved FWM between THz fields and a near-infrared probe in large band-gap (solid state) Kerr media can be employed for THz beam-profiling and imaging via standard CCD cameras. We also show an example of a space-time resolved measurement, i.e. a 3 dimensional tomographic reconstruction of the THz pulse or 3D-mapping [13].

In our experiments, we generate broadband THz pulses by focusing in air an intense (ionizing) optical field and its second harmonic, generated via a Beta Barium Borate (BBO) crystal (see Fig. 1) [14–16]. The generated THz pulse is then collimated by a 4 inch equivalent focal

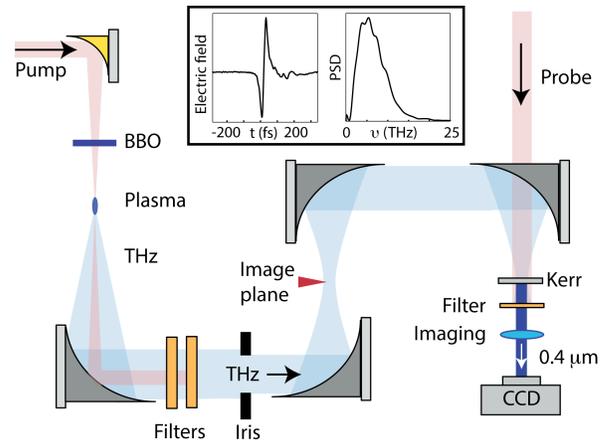

Fig. 1 (Color online). Sketch of the experimental setup employed for the generation and imaging of THz pulses. In the inset, time resolved electric field profile and spectrum recorded via ABCD detection (see text).

length–EFL gold coated parabolic mirror and the pump pulse is removed by two low-pass multi-mesh filters (QMC Instruments, UK) with 20 THz cut-in frequency and nearly 40 dB rejection for $v > 20$ THz. The THz beam is then refocused by means of a 2 inch EFL mirror to an aperture-limited spot-size. A second couple of parabolic mirrors (2 inch EFL) forms an M=1 telescope that images the focused beam (at the *image plane* in Fig.1) into the detection plane. A detailed description of the THz source and the characterization of the generated THz field can be found in [16] while in the inset of Fig. 1 we show the THz electric field and spectrum recorded via air-biased coherent detection (ABCD) [12].

In order to prove the THz imaging capabilities of FWM we employed a 500 µm thick, 4 x 4 mm aperture, <100>-cut, single crystal diamond sample as nonlinear medium and we overlapped the THz field with a collimated, 795 nm, 60 fs duration probe pulse, passing trough a 1.5 mm diameter hole in the last parabolic mirror of the THz telescope (see Fig. 1). When the THz and the probe overlap, wave-mixing occurs and a signal proportional to the THz intensity is generated at a wavelength close to the probe second harmonic (400 nm, for low conversion efficiencies $I_{WM} \propto I^2_{probe} I_{THz}$).

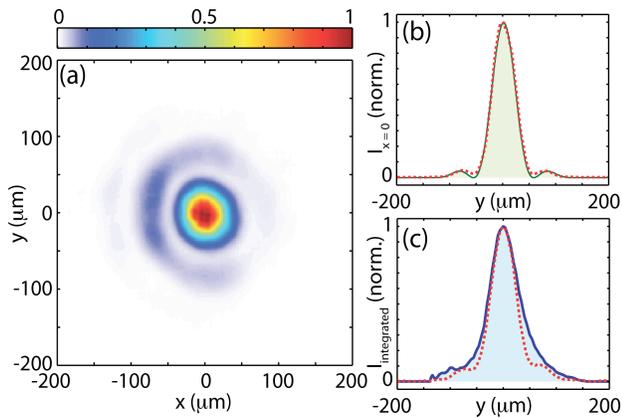

Fig. 2 (Color online). (a) THz beam profile recorded via the proposed nonlinear imaging technique. (b) Section of the THz beam along the $y$ coordinate ($x = 0$) (red, dashed curve) overlapped with a Bessel-Gauss beam fit (green, shaded). (c) Measured profile integrated on the x direction (red-dashed) compared to the result of a knife-edge measurement.

We removed the probe fundamental harmonic with a suitable band-pass filter and we imaged the diamond surface onto a cooled CCD camera (620i, QSI). In Fig. 2(a) we show the recorded image, acquired averaging over 100 laser shots, and for 350 µJ probe pulses energy. The horizontal and vertical axes are corrected in order to account for the magnification of the last imaging system. From both the imaging geometry and further optical calibrations we estimate a magnification factor M = 3.9.

The image clearly shows that the THz beam profile is Bessel-shaped, rather than Gaussian. In Fig. 1(b) we show the beam profile along the $y$-axis extracted from the image at $x = 0$ (red, dashed curve), overlapped with a fit considering a Bessel beam apodized with a Gaussian envelope, i.e. a Bessel-Gauss beam (green, shaded curve):

Eq. 1
$$I(y) = J_0^2(\alpha y)\exp\left[-(y/\sigma_y)^2\right],$$

where $\alpha \simeq 0.0412 \pm 0.0002$ µm$^{-1}$ and $\sigma_y = 76 \pm 2$ µm (we considered here the standard deviations resulting from the nonlinear fit). In order to compare the results obtained with the proposed nonlinear imaging technique with what was obtained by traditional means, we recorded the THz beam profile in the image plane both with a camera (PV 320, Electrophysics, see [16]) and with a knife-edge measurement, moving a blade in the image plane and recording the signal in the detection plane with a pyroelectric detector. The beam profile recorded with the camera and fitted by a Gaussian function resulted in $\sigma_G \simeq 85.5$ µm, compatible with the width of the Gaussian apodization recorded via the nonlinear imaging technique. We note that the large pixel size of the PV 320 camera (48.5 µm) prevents recording any fine detail, such as the Bessel structure. The THz beam-profile retrieved via the differentiation of the knife-edge data (considering an $x-y$ factorable profile) is compared in Fig. 2(c) – blue shaded

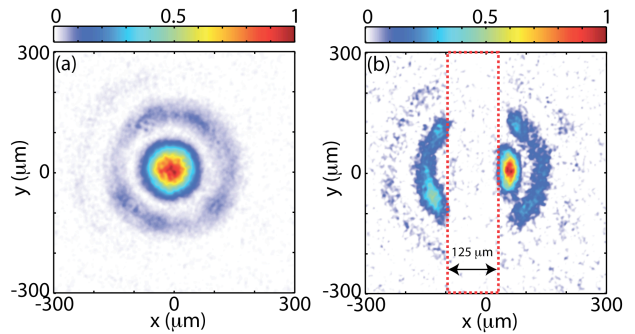

Fig. 3 (Color online). (a) THz beam profile recorded with a closed iris in the THz beam-path. (b) Same as (a) with a single mode fiber blocking the THz beam in the *image plane*.

curve – with the beam profile recorded via the proposed nonlinear technique and integrated along the $x$ direction (red, dashed curve), showing a good agreement between the two techniques. Remarkably, the beam-profile resolution of the proposed nonlinear technique is limited by the numerical aperture of the imaging system operating on the visible signal, and is hence sub-wavelength for the THz radiation.

To further prove that the Bessel profile recorded via the nonlinear technique is not an artifact, we imaged with the THz camera the beam profile for different planes along the $z$ (propagation) – coordinate, around the *image plane* (data not shown). As expected for a two-color plasma source (see e.g. [17]) we clearly observed that the THz emission is conical, with a carrier cone-angle of $\theta = 12.8 \pm 0.4$ deg (we considered the standard deviations resulting from the linear fit). This angle is indeed in agreement with the α value used in our fit (Eq. 1) assuming a carrier frequency of $\simeq 9$ THz.

To prove that the imaging capabilities of the proposed technique extend beyond simple beam profiling, we have imaged an opaque object placed in the image plane, namely a 125 µm diameter, single mode silica fiber. To this purpose we have increased the THz spot-size by closing an iris in the collimated THz beam-path (as shown in Fig. 1). In Fig. 3(a) and (b) we show the recorded visible signal without and with the fiber, respectively. As expected, Fig. 3(b) clearly shows a dark area of nearly 130 µm width, determined by the fiber.

Curiously we noted that although the THz and the probe pulse have 90 and 60 fs duration, respectively, the recorded signal is insensitive to the two-pulse delay over an interval of nearly 8 ps, as shown in Fig. 4(a). The origin of the wavelength selectivity of the proposed nonlinear imaging system, and of its insensitivity to the pulse-to-pulse delay can be found in the phase-matching properties of the FWM process employed for the THz up-frequency shifting. Indeed it has been recently demonstrated that the FWM between broadband THz and $\simeq 800$ nm pulses in diamond has a component of difference frequency generation that is naturally phase matched at $\simeq 10$ THz for fields counter-propagating to the optical pulse [18], in agreement with the 9 THz carrier frequency inferred from the above-mentioned Bessel fit. Furthermore, for a 500 µm thick diamond sample the counter-propagating wave-mixing signal is indeed expected to extend nearly 8 ps in the delay coordinate (see [18]). The image is hence generated by both the forward and, for longer delays, by the THz component reflected from the diamond-air interface (as shown in [18] this is the strongest component resulting from the FWM in diamond). On the one hand, this implies that time-resolved measurements cannot be performed with a parallel plate diamond sample. On the other hand however, the backward interaction allows to directly measuring the time-integrated beam profile with short probe pulses without scanning.

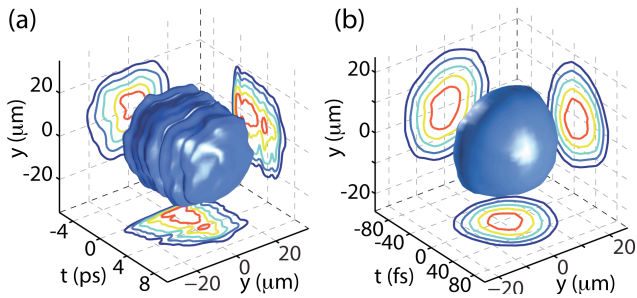

Fig. 4 (Color online). (a) and (b) are 3D-mapping of the THz pulse performed via the Kerr nonlinearity in diamond and borosilicate glass microscope coverslip, respectively. The figures shows an iso-amplitude surface at the 50% of the maximum. The curves on the sides are iso-amplitude levels for the corresponding projections (50-90%).

In order to perform space and time resolved THz characterization, it is essential to remove the backward phase matched component (due to the fact that it is delay-insensitive), which however took place in all of the transparent materials that we investigated (KBr, CsI, $Al_2O_3$, and others). An alternative to employing wedge crystals is that of relying on THz absorptive Kerr media. The THz radiation in this case is absorbed before reaching the output facet of the nonlinear medium, where the reflection occurs in the case of transparent samples. Thus the interaction occurs only in the co-propagating geometry and the measurement is sensitive to the THz-probe delay. A three dimensional mapping of the THz pulse can indeed be reconstructed employing a borosilicate glass microscope coverslip, as shown in Fig. 4(b). In this experiments the temporal duration recorded via the 3D-mapping ($\simeq$ 110 fs) is compatible with the duration of the THz pulse obtained by applying a Hilbert transform to the electric field trace recorded via air-biased coherent detection ($\simeq$ 90 fs, see inset in Fig.1) [12], once a suitable absorption spectrum at THz frequencies is considered for the borosilicate glass. From the spatial point of view, the small beam-width measured with the borosilicate glass also indicates that the mixing process converts mainly the higher frequency components of the THz pulse.

In conclusion, we have shown that imaging at THz frequencies can be performed with a standard CCD camera exploiting THz frequency up-conversion in transparent Kerr media. The counter-propagating phase-matched interaction occurring at THz frequencies allows for a delay insensitive measurement, rendering the described effect of interest as a possible alternative to focal-plane arrays operating at THz frequencies. We have also shown that space-time resolved THz mapping could be performed in absorbing Kerr media. Finally, we note that the resolution of the image obtained is not limited to the THz diffraction limit. This implies that, by performing frequency mixing directly on the object under test, sub-wavelength resolution could be achieved.

M.C. acknowledges the support from the People Programme (Marie Curie Actions) of the European Union's Seventh Framework Programme (FP7/2007-2013), REA GA 299522. D.F. acknowledges financial support from the Engineering and Physical Sciences Research Council EPSRC, Grant EP/J00443X/1 and from the European Research Council under the European Union's Seventh Framework Programme (FP7/2007-2013) / ERC GA 306559. L.C. and E.R. acknowledge the support from "Le Fonds québécois de la recherche sur la nature et les technologies." The authors acknowledge help from the ALLS staff, B.E. Schmidt, N. Thiré, M. Cassataro and M. Shalaby and support from Gentec-EO.